\documentclass[conference]{IEEEtran}
\IEEEoverridecommandlockouts
\usepackage[numbers,sort&compress]{natbib} 
\usepackage{amsmath}
\usepackage{flexisym}
\usepackage{breqn}
\usepackage[export]{adjustbox}
\usepackage{amssymb,amsfonts}
\usepackage{caption}
\usepackage{supertabular,booktabs}
\usepackage{algorithmic}
\usepackage{graphicx}
\usepackage{booktabs}
\usepackage{subcaption}
\usepackage{textcomp}
\usepackage{placeins}
\usepackage{longtable}
\usepackage{xcolor}
\def\BibTeX{{\rm B\kern-.05em{\sc i\kern-.025em b}\kern-.08em
    T\kern-.1667em\lower.7ex\hbox{E}\kern-.125emX}}
\begin{document}


%

\title{Quantifying Data Requirements for EEG Independent Component Analysis Using AMICA
\thanks{Funding provided by a gift from the Mathworks, by The Swartz Foundation (Oldfield, NY), and by the NIH grants 5R24MH120037, and 5R01NS047293.}
}

\author{\IEEEauthorblockN{Gwenevere Frank}
\IEEEauthorblockA{\textit{Electrical and Computer Engineering} \\
\textit{University of California San Diego}\\
La Jolla, USA \\
jfrank@ucsd.edu}
\and
\IEEEauthorblockN{Seyed Yahya Shirazi}
\IEEEauthorblockA{\textit{Swartz Center for Computational Neuroscience,} \\\textit{Institute for Neural Computation} \\
\textit{University of California San Diego}\\
La Jolla, USA \\
syshirazi@ucsd.edu}
\and
\IEEEauthorblockN{Jason Palmer}
\IEEEauthorblockA{\textit{Statistics} \\
\textit{West Virginia University}\\
Morgantown, USA \\
japalmer29@gmail.com}
\and
\IEEEauthorblockN{Gert Cauwenberghs}
\IEEEauthorblockA{\textit{Institute for Neural Computation} \\
\textit{University of California San Diego}\\
La Jolla, USA \\
gert@ucsd.edu}
\and
\IEEEauthorblockN{Scott Makeig}
\IEEEauthorblockA{\textit{Swartz Center for Computational Neuroscience,} \\\textit{Institute for Neural Computation} \\
\textit{University of California San Diego}\\
La Jolla, USA \\
smakeig@ucsd.edu}
\and
\IEEEauthorblockN{Arnaud Delorme}
\IEEEauthorblockA{\textit{Swartz Center for Computational Neuroscience,} \\\textit{Institute for Neural Computation} \\
\textit{University of California San Diego}\\
La Jolla, USA \\
\textit{Centre de recherche Cerveau et Cognition} \\
\textit{Paul Sabatier University}\\
Toulouse, France \\
arnodelorme@gmail.com}
}


\IEEEoverridecommandlockouts

\IEEEpubid{\makebox[\columnwidth]{978-1-6654-6819-0/22/\$31.00~\copyright2022 IEEE \hfill}
\hspace{\columnsep}\makebox[\columnwidth]{ }}

\maketitle

\IEEEpubidadjcol

\begin{abstract}
Independent Component Analysis (ICA) is an important step in EEG processing for a wide-ranging set of applications. However, ICA requires well-designed studies and data collection practices to yield optimal results. Past studies have focused on quantitative evaluation of the differences in quality produced by different ICA algorithms as well as different configurations of parameters for AMICA, a multimodal ICA algorithm that is considered the benchmark against which other algorithms are measured. Here, the effect of the data quantity versus the number of channels on decomposition quality is explored. AMICA decompositions were run on a 71 channel dataset with 13 subjects while randomly subsampling data to correspond to specific ratios of the number of frames in a dataset to the channel count. Decomposition quality was evaluated for the varying quantities of data using measures of mutual information reduction (MIR) and the near dipolarity of components. We also note that an asymptotic trend can be seen in the increase of MIR and a general increasing trend in near dipolarity with increasing data, but no definitive plateau in these metrics was observed, suggesting that the benefits of collecting additional EEG data may extend beyond common heuristic thresholds and continue to enhance decomposition quality.  
\end{abstract}


\section{Introduction}
\begin{figure}[t]
\centering
\includegraphics[width=1\linewidth]{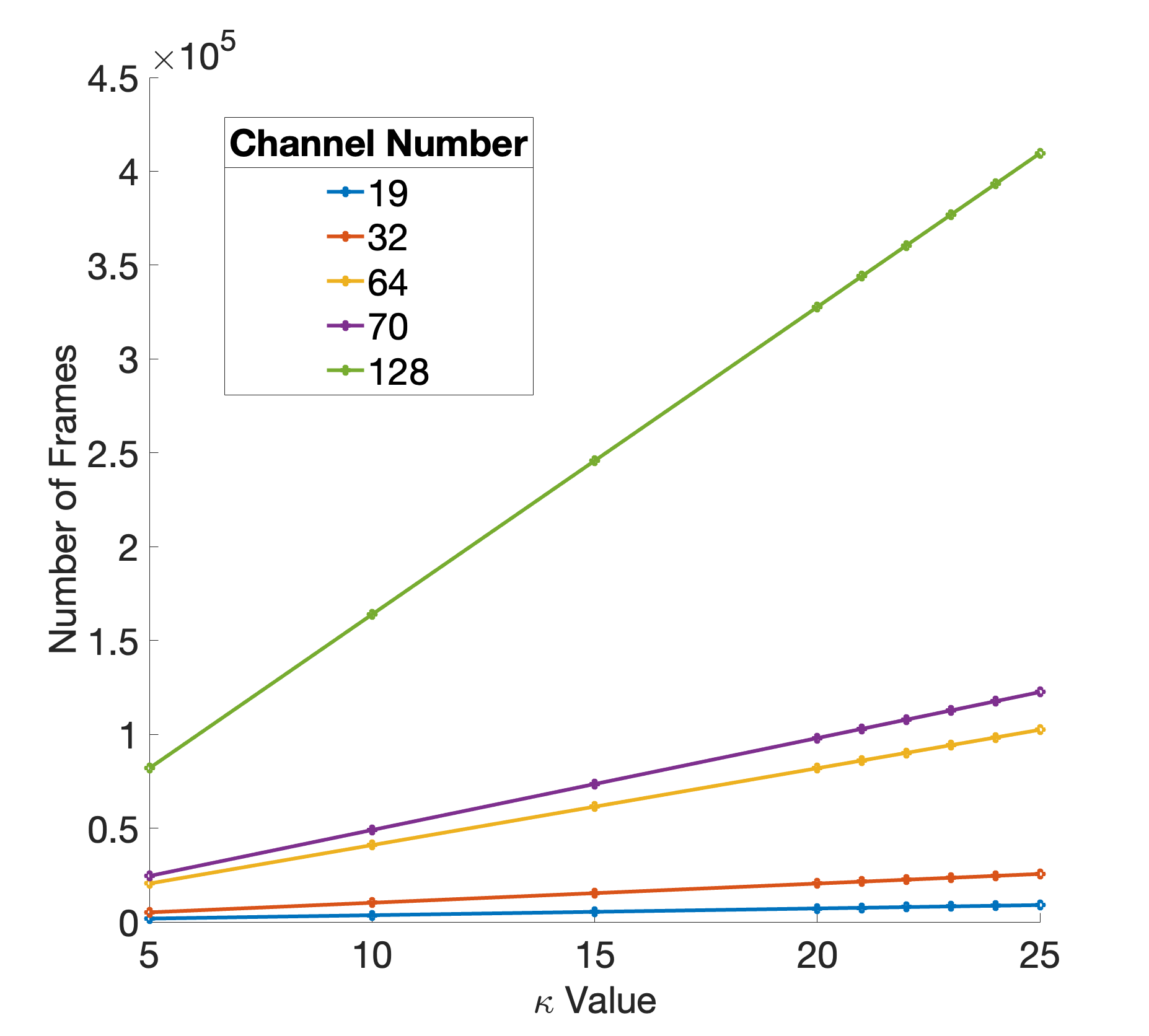}
\caption{Required number of data frames versus $\kappa$ value for multiple channel numbers}
\label{fig:kvalChanNum}
\end{figure}

Electroencephalography (EEG) captures scalp electrical potentials arising from the synchronous activity of neuronal populations. However, these recordings reflect a mixture of signals from both cortical sources and non-neural artifacts. Independent Component Analysis (ICA) is a blind source separation technique that aims to decompose these mixtures into maximally independent source signals. ICA has become a foundational method in EEG preprocessing for isolating neural activity from confounds like eye movements and muscle noise \cite{delorme2012independent}. 
EEG recordings are understood to be mixtures of synchronous field potentials originating at the cortical surface from multiple localized patches of densely connected pyramidal cells \cite{nunez1974brain,varela2001brainweb}. This understanding is supported by multiple facets of biology including the fact that neurons in close proximity share denser connections than neurons located further apart from each other\cite{stepanyants2009fractions, stettler2002lateral}, thalamocortical connections are largely radial \cite{sarnthein2005thalamocortical, dehghani2010magnetoencephalography}, and inhibitory glial networks lack long range connectivity \cite{stepanyants2009fractions}. This understanding of EEG recordings arising as mixtures of activity from local densely connected cortical patches, with the addition of non-brain signals like eye movement and powerline noise, makes independent component analysis (ICA) an ideal candidate for source separation of EEG signals. 
AMICA, or adaptive mixture ICA, is an increasingly popular ICA decomposition algorithm for EEG recordings. AMICA utilizes mixtures of generalized Gaussians to model component source densities and is capable of dividing recordings into stationary subsets and fitting a unique ICA model to each subset. AMICA has previously been shown to outperform other ICA algorithms when decomposition quality is gauged by quantitative metrics \cite{delorme2012independent, frank2022framework}. AMICA, like all ICA algorithms, requires a sufficient amount of data in order to compute high-quality decompositions. 
Data collection is often an expensive and time-consuming process, and it is important to ensure that sufficient data is collected. Because the number of parameters in the ICA unmixing matrix scales with the square of the number of channels, the number of required time points for stable decomposition becomes proportional to the number of weights to be estimated, and thus increases quadratically with the number of channels. Previously a value of 20 for the $\kappa$ value, defined as $\kappa = \frac{number\ of\ data\ frames}{(number\ of\ channels)^2}$, has been recommended heuristically \cite{delorme2012independent}. Collecting the necessary amount of data to reach a certain $\kappa$ becomes more difficult as the number of EEG channels being recorded increases, as shown in Figure \ref{fig:kvalChanNum}. The commonly used $\kappa$ value is based on heuristic practice and lacks formal empirical validation; this study aims to address that gap.

In this study, we seek to address the question: \textit{How much EEG data is required to achieve stable and high-quality ICA decompositions using AMICA?} Specifically, we hypothesize that decomposition quality—as measured by mutual information reduction and source dipolarity—will improve with increasing data quantity ($\kappa$ value), but will exhibit diminishing returns beyond a certain threshold.

\section{Methods}

\subsection{Decomposition Metrics}
Here we employ two quantitative metrics to evaluate decomposition quality.

\subsubsection{Mutual Information Reduction (MIR)}
Mutual information reduction, or MIR is the reduction in mutual information between the independent components calculated by an ICA decomposition compared to the the mutual information between channels in the time series data prior to being transformed into the independent component space. Or more precisely, MIR is a measure of the amount of mutual information removed by applying the unmixing matrix $W$ computed by ICA decomposition to the time series data $x$. In particular, since ICA attempts to minimize mutual independence between components, we can expect that an ICA decomposition that produces the most independent components will maximize MIR.

MIR also has the benefit of being able to be estimated relatively easily using one dimensional density models as described by J. Palmer in \cite{delorme2012independent}. Let $I(x)$ be the mutual information of the source components and $I(y)$ be the mutual information of the ICA components. We can then define the MIR resulting from applying $W$ to $x$ in terms of only the determinant of $W$ and the marginal entropies of x and y.

\begin{dmath}
$$ MIR = I(x) - I(y) = [h(x_1) + \dots + h(x_n)] - [h(y_1) + \dots + h(y_n)] - h(x) + log |det W| + h(x) = log |det W| + [h(x_1) + \dots + h(x_n)] - [h(y_1) + \dots + h(y_n) ] $$
\end{dmath}

A discussion of the estimation of the entropies can also be found in \cite{delorme2012independent}. In this study, MIR is reported in units of kilobits per second (Kbits/sec), reflecting the rate of mutual information reduction across the time series. This unit arises because the entropy estimates are computed in bits per sample, and when aggregated over the total sampling rate of the EEG data (250 Hz), the resulting MIR quantifies the reduction in mutual information per unit time. Thus, the reported MIR values represent the total information decorrelated by the ICA decomposition, normalized by the temporal resolution of the dataset.

\subsubsection{Dipolarity}

For each independent component of each subject, a best-fitting single dipole was calculated. These best fitting dipoles were found using EEGLAB's \cite{delorme2004eeglab} DIPFIT extension, and lacking MRI data, a simple spherical four shell boundary element method (BEM) head model was used with radii of 71, 72, 79, and 85 mm and conductances of 0.33, 0.0042, 1, 0.33 mS. A single dipole was fitted. Some components may have been better modeled by synchronous activity in two bilateral near-symmetric sources. However, these cases appear rare in this dataset. EOG activity in particular may benefit from being modeled as two near-symmetric dipoles, but with the difficulty of constructing a forward model for the front of the face and the close proximity of the two eyes together, modeling the corresponding components as a single dipole is unlikely to be a significant source of error. While modeling each component with a single dipole in a spherical model is a simplification and not realistic, this approach is commonly used for ICA quality assessment and has been shown in previous studies to provide meaningful estimates of source separability and biological plausibility \cite{delorme2012independent}. In particular, we used the metric of the percentage of sources that can be modeled by a single dipole source with a residual variance of less than 10 as a metric for ICA decomposition quality.

\begin{figure*}[t]

\includegraphics[width=1\linewidth,left]{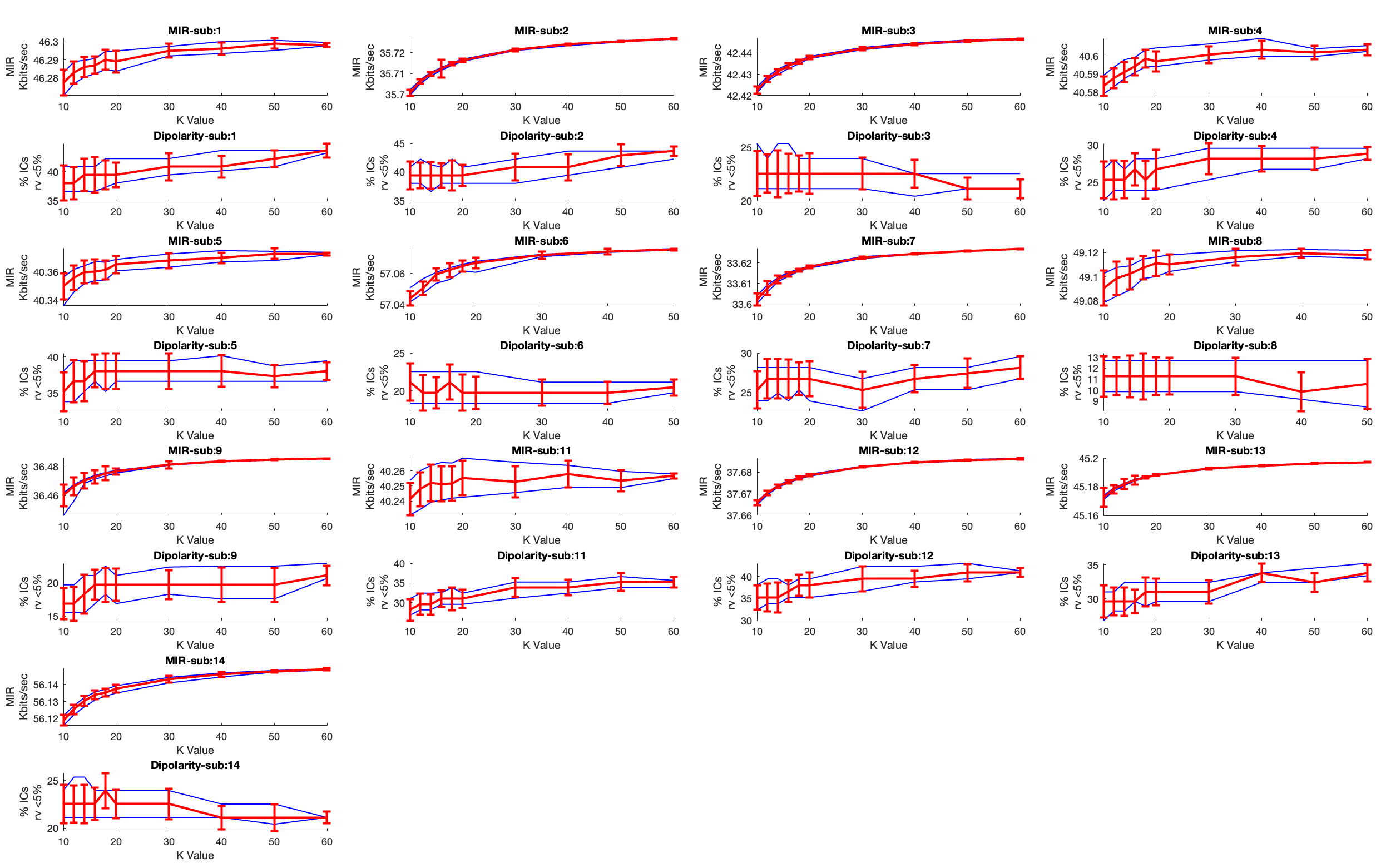}
\caption{Subplots show either MIR or the percentage of components with a residual variance of less than $5 \%$ with the best fit single dipole source plotted against multiple $\kappa$ values for each subject. The median for each dataset across random trials is plotted as a thick red line with error bars representing a single standard deviation of the data across trials. The blue lines plotted represent the 10th and 90th percentile of the MIR or near dipolarity measures across all random iterations for a given $\kappa$ value. }
\label{fig:plot1}
\end{figure*}

\subsection{The AMICA algorithm}

AMICA in past studies has been the best performing ICA algorithm when compared to other modern techniques when evaluated using quantitative metrics \cite{delorme2012independent,frank2022framework}. AMICA uses mixtures of multiple gaussian scale mixture models to estimate each source density model. AMICA also uses the Amari newton method for fast convergence on large eeg datasets. Additionally, AMICA can learn stationary subsets of data and fit ICA models for each subset of the data, however this capability was not utilized in this paper. Unless mentioned otherwise throughout this paper AMICA is run with a maximum of 3000 iterations and the rest of the parameters left at their default values.

\subsection{EEG Data}
A visual working memory task was selected for this paper, this task and dataset are common to previous works published on this subject \cite{delorme2012independent,frank2022framework,frank2023exploration}. Participants were asked to fixate on a central fixation cross for five seconds. A sequence of single letters was presented at 1 Hz. Black letters were to be memorized and green letters ignored. After the sequence, a probe letter was presented, and participants were asked to identify if the probe letter was in the memorized set. 400 ms later an auditory feedback was given. Each participant completed 100-150 trials.
Recordings contained 71 channels, sampled at 250 Hz, with each channel having less than 5 Kohm impedance. 
\subsection{EEG Data Preprocessing}
Channels were analog bandpass filtered from 0.1–100 Hz by the EEG amplifier, and the right mastoid was used as a reference. Post-collection EEG data was high-pass filtered with a 0.5 Hz linear FIR filter, epochs were extracted from -700 to +700 ms from letter onset, and the mean of each epoch was subtracted. Noisy epochs were rejected by visual inspection (between 1 and 16 per subject). Each dataset contained between 269,000–315,000 time samples. Subjects 8 and 6 have less time samples than the other datasets, leading to those datasets having a $\kappa$ value of below 60. Following our previous report, 14 subjects were selected  \cite{delorme2012independent}, 7 for having good decomposition quality based upon visual inspection of scalp topographies and 7 for having bad decomposition quality based upon the same criteria. As indicated in the previous report \cite{delorme2012independent}, one of the subject has corrupted data leaving 13 subjects. The study was conducted under IRB approval from the University of California, San Diego.

\section{Results}
\begin{figure*}[t]
\centering
\includegraphics[width=1\linewidth]{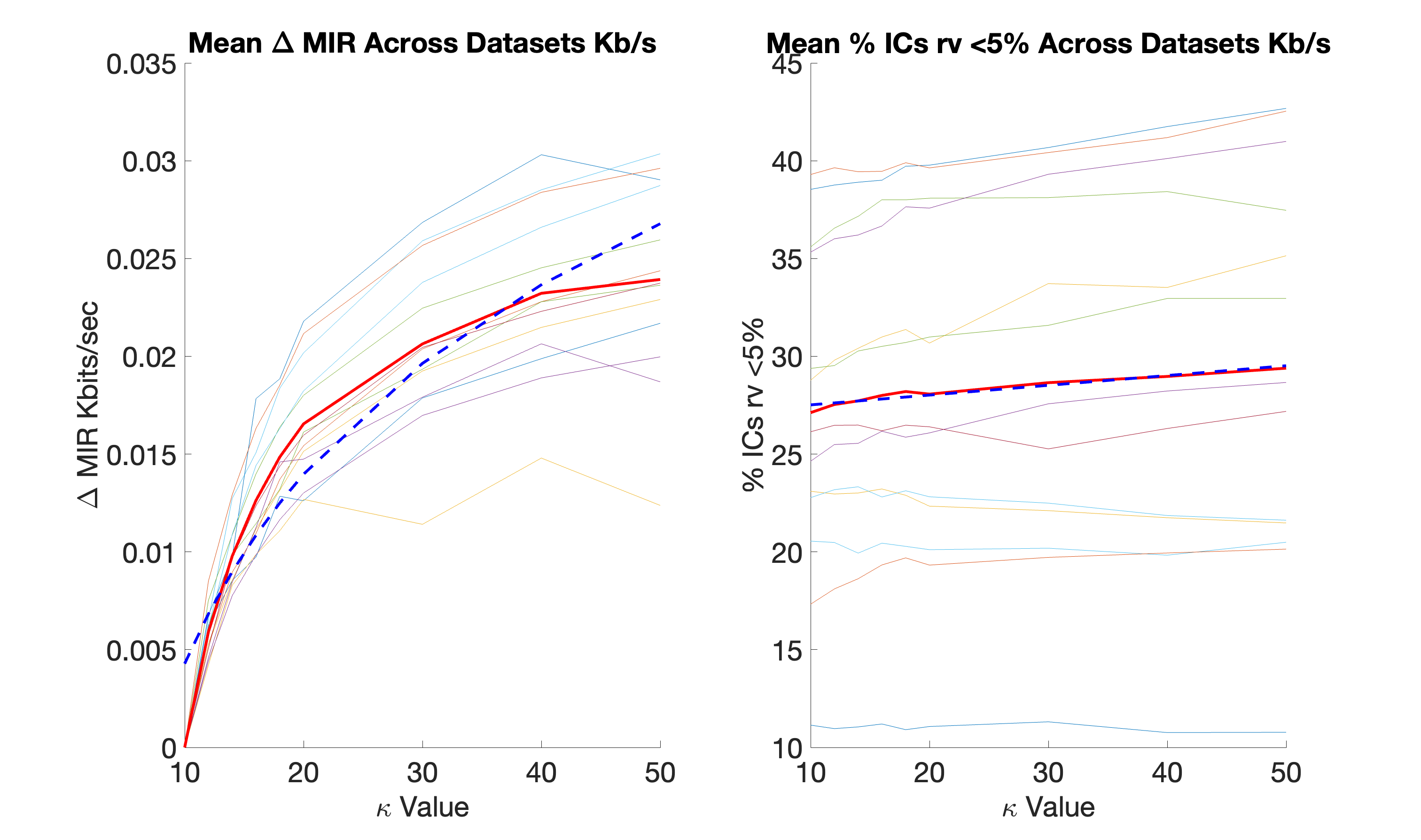}
\caption{The mean relative MIR and dipolarity across all trials is plotted against multiple $\kappa$ values for each dataset and the bold line is mean across datasets at each $\kappa$ value. Dashed blue lines are a logarithmic regression with a root mean square error of $0.00252$  for MIR and a linear regression with $R^2=0.923,P=3.72 \times10^{-5}$ for dipolarity respectively. To facilitate comparison across subjects with different baseline levels, MIR values were zero-centered by subtracting each subject’s value at $\kappa = 10$.}
\label{fig:mom}
\end{figure*}

For each dataset, multiple ICA decompositions with different random seeds -- leading to different random selections of timepoints -- were conducted for each $\kappa$ value. For each $\kappa$ value the number of decompositions with random seeds were selected so that for each $\kappa$ value across random trials approximately the same total amount of data was processed. Specifically, the number of random trials is selected as $floor(\frac{Q}{\#Channels^2 \times \kappa})$ where $Q=100 \times \#Channels^2 \times 10$ so that 100 random trials are selected for $\kappa = 10$. For each $\kappa$ value the timeframes were randomly permuted and the minimum number of timeframes required to produce a total number of timeframes over the $\kappa$ value were randomly selected from the original recordings of the subject to artificially compose a smaller dataset with a lower $\kappa$ value. Mutual information reduction and near dipolarity were then computed for each $\kappa$ value for each of the randomly sampled datasets for each of the 13 subjects. Near dipolarity was computed by fitting the best equivalent dipoles to each independent component and computing the number of components with a residual variance less than five to their best fit equivalent dipoles. Median MIR and near dipolarity across random samplings were plotted for each subject. Figure \ref{fig:plot1} shows the results for the 13 subjects. Moreover, Figure \ref{fig:mom} also shows the mean MIR for each subject plotted against each other with the mean across all the subjects and random samplings plotted in bold in the forefront, and an equivalent figure for dipolarity.

For each subject the median MIR fluctuates less than 31 bits/sec across all the tested $\kappa$ values. For each subject the median near dipolarity fluctuates less than 7\% across all the tested $\kappa$ values. The mean MIR across all random samples and subjects shows a generally increasing trend up to $\kappa=50$. The median MIR for each subject shows a generally increasing value up to $\kappa = 60$ (or $\kappa=50$ for datasets 6 and 8), with some noise, for all subjects. The mean near dipolarity across all random samples and subjects also shows a generally increasing trend up to $\kappa=50$. The median near dipolarity for each subject also shows a generally increasing value with increasing $\kappa$ with the exclusion of subjects 3, 6, 8, and 14.

\begin{figure}[t]
\centering
\includegraphics[width=1\linewidth]{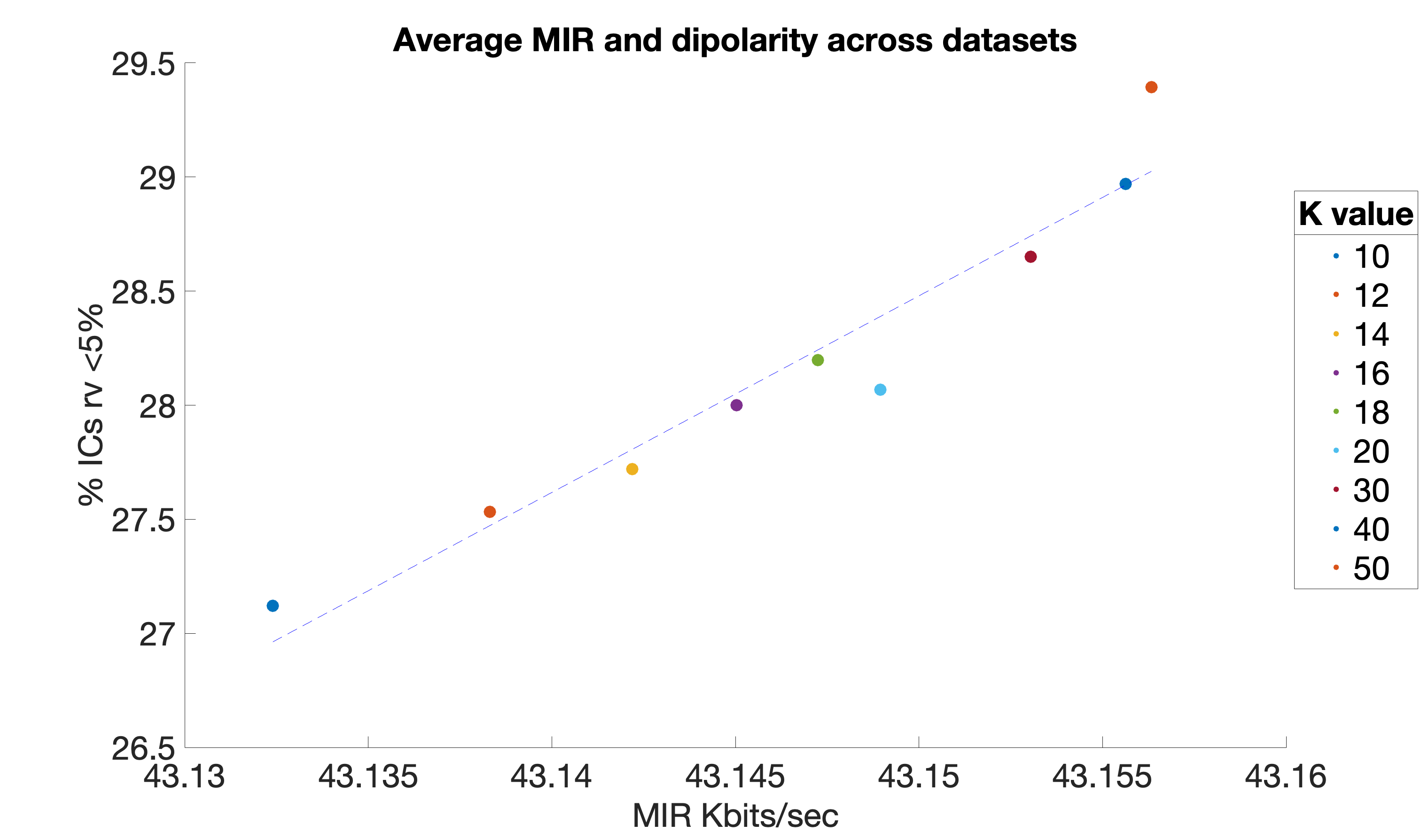}
\caption{The mean MIR and dipolarity values across random samplings and subjects for various $\kappa$ values. The blue line is a linear regression with $R^2=0.3590, P=0.0672$ }
\label{fig:scatter}
\end{figure}

Finally, Figure \ref{fig:scatter} shows the median MIR and near dipolarity averaged across all random samplings and all subjects for each $\kappa$ value. It can be seen that the MIR/near dipolarity relationship follows a linear trend with increasing $\kappa$ ($R^2=0.3590, P=0.0672$).

Note that for Figure \ref{fig:mom} and Figure \ref{fig:scatter} since all subjects are being compared a maximum $\kappa$ value of 50 is used so as to include subjects 6 and 8.

\section{Discussion}

The results show an increasing trend in mutual information reduction with a clear asymptotic behavior for subjects 2, 3, 6, 7, 9, 12, 13, and 14 as $\kappa$ increases up to $\kappa=50$ and a general increase in near dipolarity with higher $\kappa$ values. However, no definitive plateau is observed across the tested range. Despite the relatively large number of data points per subject (approximately 300,000), this was not sufficient to rigorously test whether decomposition quality saturates beyond a certain $\kappa$ threshold. Nevertheless, the observed trend supports the view that increasing data quantity leads to improved ICA decomposition, with greater source independence and dipolarity—both indicators of increased biological plausibility. Future studies with larger datasets will be needed to determine whether a point of diminishing returns can be reliably identified.

We also reproduce the linear relationship between mutual information reduction and dipolarity observed in previous studies, notably by Delorme et al. (2012) \cite{delorme2012independent}, though in a different context—here, the trend arises within a single ICA algorithm (AMICA) as a function of data quantity, rather than across multiple ICA algorithms. This finding is particularly striking given that MIR and dipolarity measure fundamentally different aspects of decomposition quality. MIR quantifies statistical independence among components, based on entropy and mutual information, while dipolarity reflects the spatial plausibility of components as estimated neural sources, derived from fitting equivalent dipoles. That these two unrelated metrics exhibit a strong linear association across varying data quantities suggests the possibility of an intrinsic link between statistical independence and the emergence of spatially compact, physiologically plausible sources in EEG. This convergence is not guaranteed by the ICA objective function and may reflect deeper properties of the cortical source structure and the mixing process in scalp EEG. Further work is needed to understand the conditions under which this relationship holds and whether it generalizes across tasks, datasets, and ICA implementations.

AMICA was selected for this study because prior work has shown it to outperform other commonly used ICA algorithms—such as Infomax and FastICA—when evaluated using objective metrics like mutual information reduction and source dipolarity \cite{delorme2012independent}. Given these results, AMICA was deemed the most appropriate choice for exploring the effects of data quantity on decomposition quality. While a side-by-side comparison with alternative algorithms would provide useful context, the focus of this study was on quantifying data sufficiency using the best available method, rather than benchmarking algorithm performance. In addition, the choice of parameters for running AMICA, particularly the maximum number of iterations, naturally raises the question of whether the algorithm consistently reached convergence and whether higher iteration limits might yield better decompositions. In this study, AMICA was run with a maximum of 3000 iterations and default settings for all other parameters. To assess the adequacy of this choice, we performed additional simulations on a representative subset of datasets with an increased limit of 5000 iterations. These extended runs showed no meaningful improvement in decomposition quality, as evaluated by mutual information reduction and dipolarity. This indicates that 3000 iterations were sufficient for convergence under our experimental conditions, and we observed no significant variation in performance across subjects attributable to iteration limits.

Subject-level variability is a well-known characteristic of ICA decomposition quality in EEG, and in our study, certain subjects such as subject 10 exhibited outlier behavior. While we did not perform a dedicated analysis of these deviations, we acknowledge their importance and note that understanding the sources of such variability is a valuable direction for future work. The reasons for this variability remain uncertain, but plausible contributing factors include differences in signal quality, variability in cortical source orientation, or a higher prevalence of muscle or other non-brain artifacts in some subjects. A more comprehensive analysis across a larger cohort would be needed to systematically assess these possibilities and will be considered in future investigations.

\section{Conclusion}
The analysis and results presented in this study offer EEG researchers a more rigorous, quantitatively grounded understanding of the relationship between data quantity and ICA decomposition quality using AMICA. The consistent increase in mutual information reduction and dipolarity metrics with higher $\kappa$ values supports the recommendation to collect as much data as is practically feasible to enhance the quality and biological plausibility of decompositions. No definitive plateau in these metrics was observed across the tested range of $\kappa$ values, suggesting that the benefits of additional data may extend beyond current heuristic thresholds. Future work with datasets featuring even higher $\kappa$ values will be necessary to determine whether decomposition quality ultimately saturates and to identify any potential upper bounds for optimal data collection.

\FloatBarrier

\bibliographystyle{IEEEtran}
\bibliography{IEEEabrv,refs}

\end{document}